# Dynamic interferometry measurement of orbital angular momentum of light

**Hailong Zhou, Lei Shi, Xinliang Zhang, and Jianji Dong[*]**

*Wuhan National Laboratory for Optoelectronics, Huazhong University of Science and Technology, Wuhan, 430074, China*

*Corresponding author: jjdong@mail.hust.edu.cn

We present a dynamic interferometry to measure the orbital angular momentum (OAM) of beams. An opaque screen with two air slits is employed, which can be regarded as the Young's double-pinhole interference. When the OAM beams with an annular intensity distribution vertically incident, the far-field interference patterns depend on the phase difference of the light in the two pinholes. We scan the angle between the two slits, the output intensity at center changes alternatively between darkness and brightness. Utilizing this characteristic, we can measure the OAM of light. This scheme is very simple and low-cost. In addition, it can measure very large topological charge of OAM beams due to the continuously scanning.

Light beams carrying orbital angular momentum (OAM) are associated with an azimuthal phase structure $\exp(il\theta)$, where $\theta$ is the angular coordinate and $l$ is the azimuthal index, defining the topological charge (TC) of the OAM beams [1]. These beams have an OAM of $l\hbar$ per photon ($\hbar$ is Planck's constant $\hbar$ divided by $2\pi$). In recent years, OAM beams have been widely used in a variety of interesting applications, such as optical microscopy [2], micromanipulation [3,4], quantum information [5,6], free-space and fiber optical communication [7-9].

The TC of OAM beams characterizes the corresponding mode [7-9] and the magnitude of optical torque [10,11], so the capability of distinguishing different OAM states is very important. There are several existing methods to detect optical vortices, such as interference with a plane wave [12], self-homodyne detection [13], Cartesian to log-polar coordinate transformation [14,15], diffraction patterns of various apertures or slits [16-25]. Interference with a plane wave is the most common way to measure the TC of OAM beams in the laboratory but needs an additional reference beam. Self-homodyne detection converts the TC to the voltage and is convenient to process by computer or other digital processing system. While it employed $90^0$ or $180^0$ hybrids, it is inconvenience to couple beams to waveguide and the operation bandwidth is restricted by hybrids. In addition, it is only applicable for high pure OAM beams with a small TC because only two sampling points are employed. Cartesian to log-polar coordinate transformation can detect multiple modes at the same time. Nonetheless, The method has a limitation due to the overlap of the spots for different OAM states Diffraction patterns of various apertures or slits such as square and triangular slits [21-23] can also be used to measure the TC but only for small TC (<20 and <10 for square and triangular slits respectively). Double-slit interference [24,25] does not need additional reference light and is a linear model. In our previous report, we integrated the function of polarization filter into the device. However, this model strongly relies on the fringe resolution, so it is still difficult to distinguish large TC. The phenomenon of the other methods [16-20] is not intuitional or regular, so the patterns of different OAM beams are difficult to identify.

In this letter, we put forward a dynamic interferometry to measure OAM of beams. An opaque screen with two air slits is employed, which can be regarded as the Young's double-pinhole interference. When the OAM beams with an annular intensity distribution vertically irradiate the screen, the far-field interference patterns depend on the phase difference of the light in the two pinholes. The intensity is converted to the voltage by a photoelectric detector (PD) and it is convenient to process by digital processing system. We scan the angle between the two slits, the output intensity at center changes alternatively between darkness and brightness. Utilizing this characteristic, we can measure the OAM of light. This scheme is very simple and low-cost. In addition, it can measure very large TC of OAM beams due to the continuous scanning. It also has a quite large tolerance for central deviation and mode impurity.

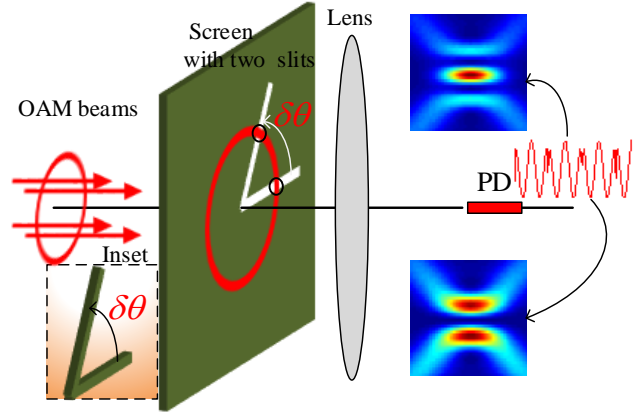

FIG. 1. Schematic structure of the dynamic interferometry with double slits. Inset, equivalent double arms.

Figure 1 presents the schematic structure of the proposed setup. The setup is composed by an opaque screen with two air slits, a Fourier transform lens and a PD. When the OAM beams with an annular intensity distribution vertically incident on the screen, only two pinholes have light transmitted shown as the black circles in Fig. 1. The lens is used to shift the interference patterns to center by Fourier transform, then the PD converts the intensity to the voltage. The voltage is minimum when the interference is destructive and is maximum when the interference is constructive. Assume the waves in the two pinholes are plane with different phase and the distributions of amplitudes are $f(x, y)$ around their own center. We define that

$$\mathscr{F}[f(x, y)] = F(u, v), \tag{1}$$

where $\mathscr{F}[\cdot]$ denotes the spatial Fourier transform operation only on the transverse coordinates $(x, y)$ and $(u, v)$ is the coordinate of spatial frequencies. So the two interferential waves in the receiving plane $(x_2, y_2)$ can be wrote as

$$\begin{aligned} E_i(x_2, y_2) &= \mathscr{F}[f(x - R\cos\theta_i, y - R\sin\theta_i)\exp(jl\theta_i)] \\ &= F(u, v)\exp(jl\theta_i)\exp[-j2\pi R(u\cos\theta_i + v\sin\theta_i)] \end{aligned}, \tag{2}$$

where $R$ is the radius of the OAM beams in the screen and $\theta_i (i = 1, 2)$ is the azimuth angle of the two slits (the angle $\delta\theta$ between the slits equals $\theta_2 - \theta_1$). $x_2 = u\lambda f$, $y_2 = v\lambda f$ where $\lambda$ is the input wavelength and $f$ is the focus length of the lens. Near the optical axis ($(x_2, y_2) \approx (0, 0)$) the interferential waves is simplified by

$$E_i(x_2, y_2) \approx F(u, v)\exp(jl\theta_i). \tag{3}$$

So the interferential intensity distribution is

$$I(x_2, y_2) = |E_1 + E_2|^2 \propto (1 + \cos l\delta\theta). \tag{4}$$

From Eq. (4), one can see that the intensity distribution has a relationship of cosine with the TC multiplying the angle between the slits. And the electrical signal from the PD will be proportional to the received power. So we can determine the TC of the input OAM beams by scanning the angle.

In the following, we will prove the aforementioned derivation by scalar diffraction theory [26]. The wave at the receiving plane is the Fraunhofer diffraction of the one behind the screen, so the intensity can expresses as

$$I(x_2, y_2) \propto \left| \iint E_{in}(x, y) P(x, y) \exp[\frac{-j2\pi}{\lambda f}(xx_2 + yy_2)] dxdy \right|^2, \quad (5)$$

where $E_{in}$ is the input OAM wave and $P(x, y)$ is the transmittance function of the opaque screen with two air slits. We set $\theta_1 = 0$ and assume the input beam is an ordinary Gaussian vortex beam, expressed as [27]

$$U_{in1}(r, \theta) = (r/w)^{|l|} \exp(-r^2/w^2) \exp(il\theta), \quad (6)$$

where $(r, \theta)$ are two-dimensional polar coordinates corresponding to rectangular coordinates $(x, y)$ and $w$ is the waist size. The radius of the OAM beams can be deduced that $R = w\sqrt{(l/2)}$.

We set the wavelength of input beams as 1550 nm. The waist size equals 5 mm, the focal length of lens is 200 mm and the width of slits is $d = 1$ mm. Figure 2 presents the dependence of the received power of the PD on the angle between the slits. From Fig. 2(a), we can see the receive signals are periodic and the period equals $2\pi$. The number of peaks or troughs in a period equals the TC of input OAM beams. The results agree well with the theoretical derivation by Eq. (4) except the first split peak. This phenomenon can be easily understood. When the angle is near $2\pi m$ ($m$ is an arbitrary integer), the two slits overlap so that the transmitted light energy decreases. In the most extreme case when the angle equals $2\pi m$, there is only one output wave and the energy is approximately equal to the quarter of the other peak power. Thus this peak will split into two. This characteristic is very beneficial to distinguish the period. To clarify more clearly, Fig. 2(b) shows the normalized power in polar coordinates, the number of petals represent the corresponding OAM beams and the first one splits into two owing to the overlap of the two slits. Other examples are also given when TCs equal $\pm 1, \pm 2, \pm 3, \pm 10$, as shown in Fig. 2(c).

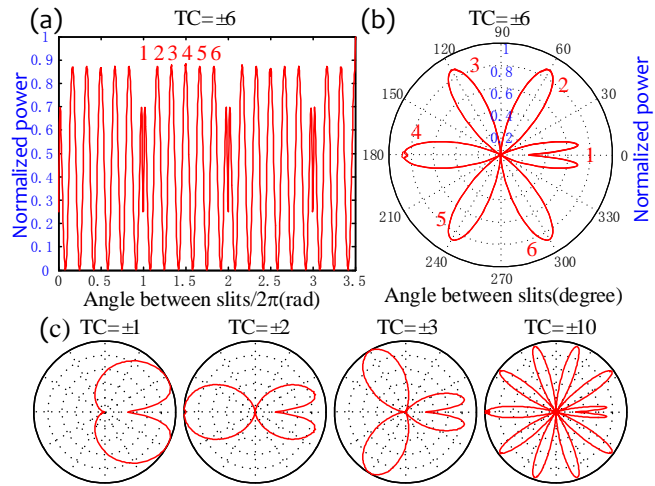

FIG. 2. Received power dependence on the slit angle in (a) Cartesian and (b-c) polar coordinates when different OAM beams vertically irradiate.

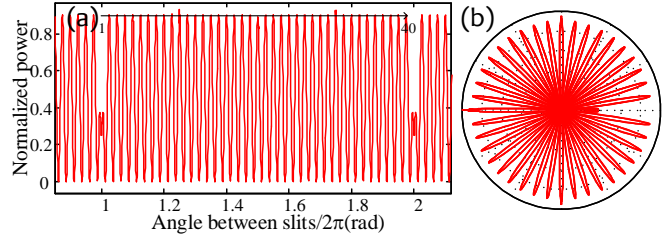

FIG. 3. Received power when TC equals ± 40. (a) Cartesian coordinate, (b) polar coordinate

The main advantage of the dynamic interferometry measurement is its ability to detect OAM mode with a very large TC benefited from the continuously scanning and it is convenient to process by digital processing system. For example, when an OAM beam with a TC equal to ± 40 vertically incidents, the received signals are still precise as shown in Fig. 3(a-b).

The main factor which limits the magnitude of detected OAM is the width of slits. When the width is too large to sample the $2\pi$ changes of phase, it will miss some peaks, thus some key information will be lost. For example, when setting $d$ = 15 mm and TC = ± 40, Figure 4(a) presents the received signals. One can see that the signals deteriorate and only 38 peaks appear. When keeping increasing the slit width or the TC of OAM, the number of lost peaks will increase because the slits cannot sample the phase precisely. But the slit width of 1 mm is precise enough for most applications.

This scheme also has a quite large tolerance for central deviation and mode impurity. Figure 4(b) presents the output signals when TC = ± 6 and the central deviation is half of radius (4.3 mm). Even so, the scanning power signal exhibits a TC of 6 or -6 regardless of the peak distortion and different petal widths. The distortion is caused by the nonuniform phase distributions around the center axis. As shown in Fig. 4(b), the hot picture displays the phase of incoming OAM beam in the screen, where the red circle represents the location of the maximum intensity and the black lines indicate the two slits. The OAM beam is obviously off-centered. When the angle of the second slit scans from 0 to $2\pi$, the changing rate of phase in left side is slower than the one in right side. So the number of peaks in left side is less. The key element of this scheme is to count how many times the phase changes by $2\pi$, so it has a tolerance for mode impurity as long as the mode impurity does not influence the times of phase change by $2\pi$. For example, Fig. 4(c) shows the received power when the purity of TC = 6 is 80% and the ones of the other two modes (TC =7, 5) are both 10%. The width and the amplitude of received power distribution are changed by other OAM modes, but we can still infer the main mode is the one whose TC equals 6 or -6.

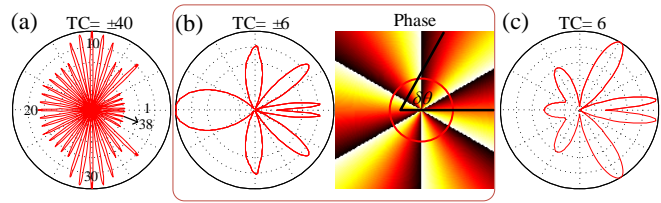

FIG. 4. Received power when (a) $d$ = 15 mm. (b) Shift 0.5 $R$. (c) Interferometry measurement of impure mode superposition with 80% $OAM_6$, 10% $OAM_7$, and 10% $OAM_5$.

All analysis above is not relevant to the sign of TCs. In fact, it fails to distinguish the sign in current approach. This characteristic can be also inferred from Eq. (4) where the cosine function is even symmetric. To distinguish the sign of TC, we must break the even symmetric. It can be realized by tinily tilted light incidence. Assume that the OAM beam tilted irradiates in $(x_2, z)$ plane. In this case, the cosine term in Eq. (4) turns into $\cos(l\delta\theta + k_x x_2)$ ($k_x$ is a constant). Then Eq. (4) turns to be relevant to the sign of TC. Figure 5 presents the phase in the screen and the received power when the OAM beam tinily tilted irradiates with a TC equal to 6 or -6. The black circle represents the location of the maximum intensity. It can be seen that the phase distributions are quite uneven and are related to the sign, that is, the phase change in upper half plane is slower than the one in lower half plane when TC > 0 (Fig. 5(a)) and is inverse when TC < 0 (Fig. 5(b)). So the received power of the two modes is exactly inverse when scanning the angle. One can see that the power petal is wider when the corresponding phase change is slower. Although the tilted incidence can detect the sign of TC, the measured magnitude of TC is prone to be wrong because the phase distribution in the screen is sensitive to the incidence angle. A little incline of incidence will add a tightly phase grating to the OAM beams and seriously break the helical phase structure. Thus the received power will be chaotic since the helical phase structure is broken, and it is hard to sample the fast phase variation accurately. As shown in Fig. 6(a), the times of phase change in $2\pi$ along the maximum intensity (black circle) become a very large value when the OAM beams with TC = 6 incident with a little tilted angle, which will cause an erroneous measurement results shown in Fig. 6(b). So we must precisely tune the incidence angle, or only use the tinily tilted incidence to determine the sign and measure the magnitude accurately by vertically incidence.

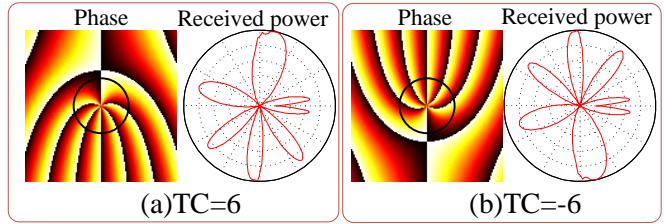

FIG. 5. Phase in the screen and received power when tinily tilted incidence.

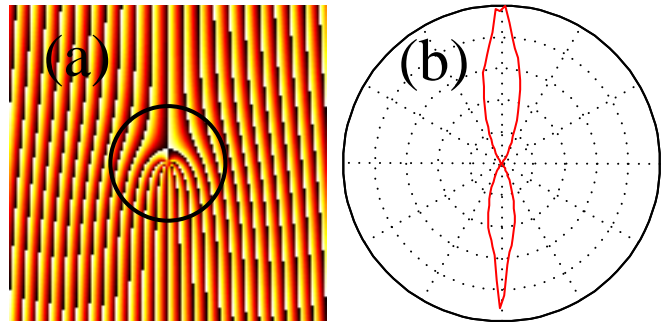

FIG. 6. (a) Phase and (b) received power of TC = 6 in the screen under a little tilted angle incidence.

The main difficulty of the proposed scheme is how to achieve the scanning of the angle between two slits. It can be realized by using spatial amplitude modulation devices such as binary-amplitude spatial light modulator [28,29]. Here we suggest a simple and equivalent method. The equivalent structure is shown in the inset of Fig. 1. The double-arm structure is complementary to the double-slit structure, it means the sum of the transmittance functions of the two structures is 1. Since there is no intensity at center for OAM beams, the received intensity in this case can expresses as

$$I(x_2, y_2) \propto \left| \iint E_{in}(x, y)[1 - P(x, y)] \exp[\frac{-j2\pi}{\lambda f}(xx_2 + yy_2)] dxdy \right|^2$$
$$\approx \left| \iint E_{in}(x, y) P(x, y) \exp[\frac{-j2\pi}{\lambda f}(xx_2 + yy_2)] dxdy \right|^2_{(x_2, y_2) \approx (0,0)} . \quad (7)$$

From Eq. (5) and Eq. (7), it proves the two structures have the same function and the simulations also indicate the received power of the two structures is almost the same. In addition, the double-arm structure is easy to fabricate and easy for scanning.

In conclusion, we put forward a dynamic interferometry to measure OAM beams. An opaque screen with two air slits is employed, which can be regarded as the Young's double-pinhole interference. When the OAM beam with an annular intensity distribution vertically incidents, the far-field interference patterns depend on the phase difference of the light in the two pinholes. The intensity is converted to the voltage by a PD and it is convenient to process by digital processing system. We scan the angle between the two slits, the output intensity at center changes alternatively between darkness and brightness. Utilizing this characteristic, we can measure the OAM of light. This scheme is very simple and low-cost. In addition, it can measure very large TC of OAM beams due to the continuously scanning. This scheme also has a quite large tolerance for central deviation and mode impurity. Besides, our scheme can be extended to measure any engineering structured light which is characterized by azimuthal phase distribution, such as optical vortices and so forth.

This work was partially supported by the Program for New Century Excellent Talents in Ministry of Education of China (Grant No. NCET-11-0168), a Foundation for the Author of National Excellent Doctoral Dissertation of China (Grant No. 201139), the National Natural Science Foundation of China (Grant No. 11174096) and the Fundamental Research Funds for the Central Universities, HUST: 2014YQ015.